\documentclass[12pt,a4paper]{article}

\usepackage{amsmath}
\usepackage{amsthm}
\usepackage{amscd}
\usepackage{amsxtra}
\usepackage{upref}
\usepackage{amsfonts}
\usepackage{amssymb}
\usepackage{graphicx}
\usepackage{stmaryrd}
\usepackage{url}
\usepackage{cite}
\usepackage{pifont}

\def\beq{\begin{equation}}
\def\eeq{\end{equation}}

\newcommand{\WI}{\left(W^{-1}\right)}

\newcommand{\ph}{\ensuremath{\phantom}}

\newcommand{\ssapprox}{ \,\hat{\approx}\,}

\theoremstyle{definition}

\numberwithin{equation}{section}

\begin{document}

\title{The $O_{D,D}$ Geometry of String Theory}
\author{David S. Berman.\\ School of Physics and Astronomy,\\ Queen Mary University of London,\\ Mile End Road,\\
London E1 4NS,\\ England\\ \\ \\ \\  Chris D. A. Blair, Emanuel Malek and Malcolm J. Perry,
\\DAMTP,\\Centre for Mathematical Sciences,\\Wilberforce Road,\\Cambridge CB3 0WA,\\
England}
\maketitle
\begin{abstract}
We construct an action for double field theory using a metric connection that is compatible with both the generalised metric and the $O_{D,D}$ structure. The connection is simultaneously torsionful and flat. Using this connection one may construct a proper covariant derivative for double field theory. We then write the doubled action in terms of the generalised torsion of this connection. This action then exactly reproduces that required for double field theory and gauged supergravity.
\end{abstract}
\section{Introduction}

The NSNS sector of the closed superstring contains massless excitations that have the quantum numbers of a graviton, an abelian $2$-form potential $B_{mn}$, usually termed the Kalb-Ramond field, and a dilaton $\Phi$. These excitations can form classical condensates that are the metric of spacetime $g_{mn}$ together with the other fields propagating in it. Conformal invariance of the string gives the equations of motion for the dilaton and Kalb-Ramond fields coupled to gravity, which can be derived from the action ~\cite{Callan:1985ia}
\beq
S = \int d^{10}x~ \sqrt{-g}~ e^{-2\Phi}\Bigl(R - \frac{1}{12}H_{mnp}H^{mnp}-4\nabla_m\Phi\nabla^m\Phi
+4\Box\Phi\Bigr) \,,
\label{eq:action}
\eeq
where $R$ is the Ricci scalar of the metric $g_{mn}$
and the field strength $H_{mnp}$ of the $2$-form potential $B_{mn}$ is
\beq
H_{mnp}=3\partial_{[m}B_{np]}\, .
\eeq

If some directions of space are compactified to give a $D$-dimensional torus, then the spacetime is no longer simply connected and strings can wind around the compact
directions. String duality means that these windings can be reinterpreted as momenta in a dual space. Similarly momenta in the original space are interpreted as windings
in the dual space. If these spatial directions are associated with a circle isometry, then it is possible to construct an explicit map from one space into the other. This map is what is usually known as T-duality. The T-duality group is $O_{D,D}$ and it maps various physical $D$-dimensional spacetimes into each other. These different spacetimes could be regarded as separate objects. One is however led to investigate a formulation of the string in which both the space and its T-dual are simultaneously present. In this formulation T-duality is no longer a hidden symmetry. The $D$-dimensional spacetime is replaced by a $2D$-dimensional space, together with a rule for selecting out which of these dimensions will be regarded as the physical $D$-dimensional spacetime. T-duality transformations amount to picking out different $D$-dimensional spacetimes from the $2D$-dimensional object. The role of $O_{D,D}$ is now that coordinates lie in its vector representation. T-duality transformations are identified with the Weyl group of $O_{D,D}$. It seems hard to believe that these doubled directions only exist if there is a spacetime isometry and so one is led to considering as fundamental an entire $2D$-dimensional doubled space.

Another motivation for searching for a more general formulation of string theory, that leads to the same endpoint, is to note that the action ~(\ref{eq:action}) is
only partially geometric. The action for the spacetime metric is of course geometric and yet the fields $B_{mn}$ and $\Phi$ are described as living in the spacetime described by the metric. It is more satisfying for all the fields in the theory to be on the same footing and so we seek a formulation where the two-form field $B_{mn}$ and
ordinary metric are combined into a single {\it{generalised geometric}} entity. The dilaton will have to remain separate in the $O_{D,D}$ description of the theory,
although when one extends this approach to make the whole U-duality group  manifest, the dilaton also becomes included in the generalised metric. The initial idea of making $O(D,D)$ a manifest symmetry was in Duff ~\cite{Duff:1989tf}i and then extensively developed by Tseytlin \cite{Tseytlin:1990nb,Tseytlin:1990va}. The double field theory (DFT) formalism was then introduced by Siegel \cite{Siegel:1993xq,Siegel:1993th}. A related approach known as generalised geometry was undertaken by Hitchin ~\cite{Hitchin:2004ut, Hitchin:2010qz} and Gualtieri \cite{Gualtieri:2003dx} where only the tangent space is extended, making the $O_{D,D}$ structure apparent on the extended tangent bundle. In DFT one doubles the entire space such that the tangent space of the doubled space is the extended tangent space of generalised geometry.

This is the doubled formulation of string theory. It has recently gone through a renaissance \cite{Hull:2009mi,Hull:2009zb,Hohm:2010jy,Hohm:2010pp}. Double field theory is actually an example of a theory that can be found from a non-linear realisation, as in the $E_{11}$ programme of West et al
\cite{West:2001as,Schnakenburg:2001he,Riccioni:2007au,Riccioni:2007ni,Riccioni:2009hi,
Riccioni:2009xr,West:2004st,Berman:2011jh,Godazgar:2013rja}. Some details of how this
construction works for doubled geometry can be found in ~\cite{Rocen:2010bk}. We will,
however, not pursue that avenue in this paper.

Significant work has been done in the generalised geometry of string and M-theory by
Waldram and collaborators, see ~\cite{Coimbra:2011nw,Coimbra:2011ky,Coimbra:2012af}. Type II double field theory was developed in \cite{Hohm:2011zr,Hohm:2011dv}
The first description complete description of doubled frame geometry was in \cite{Hohm:2010xe}.
Other interesting and relevant work on generalised geometry and double field theory can be found in ~\cite{Berman:2007xn,Berman:2007yf,Maharana:2013uvy,Dibitetto:2012rk,Andriot:2012an,Park:2013gaj,Cederwall:2013oaa,
Cederwall:2013naa,Berman:2010is,Berman:2011pe,Berman:2011kg,Malek:2012pw,Malek:2013sp,Jeon:2012hp,Jeon:2010rw,Jeon:2011kp,Hohm:2012gk,Hohm:2011cp,Hohm:2011ex,Hohm:2011nu}.

The paper is structured as follows. First, we describe in section \ref{sec:2} the metric and vielbein of generalised geometry. Following on from that in section \ref{sec:3}  we describe the local symmetries of the theory in terms of a generalised Lie derivative, and discuss what it means to be a generalised tensor. This necessitates the introduction of the idea of a {\it{weak}} and {\it{strong}} tensor. In section \ref{sec:4} we then describe the main result of the paper, which is the construction of the connection from which one can form the action for DFT. Then, in section \ref{sec:5}, we apply these ideas to a generalised Scherk-Schwarz reduction \cite{Dall'Agata:2007sr,Aldazabal:2011nj,Geissbuhler:2011mx,Grana:2012rr,Berman:2012uy,Musaev:2013rq} and show that it produces the right action for gauged supergravity. This includes terms that were previously added by hand; using the construction described in this paper those terms are shown to be a natural consequence of our formalism. We end with a discussion of the possible application of these ideas to future work.

\section{Metric and Vielbein}
\label{sec:2}

In general relativity, the metric of spacetime at each point belongs to ${\mathbb R}^+\otimes
\Sigma$ where $\Sigma$ is the symmetric space $G/H$ with $G=SL_D$ and $H=SO_{D-1,1}$
for the Lorentzian theory or for a Euclidean version of the theory, $H=SO_D$. 
The metric can be written in terms of the vielbein $e^\alpha{}_i$ by
$g_{ij}=e^\alpha{}_i e^\beta{}_j \bar\eta_{\alpha \beta}$ which shows that local $H$-transformations
are a symmetry of the spacetime metric.

In doubled geometry, the situation is very similar. The NSNS sector of the superstring
contains the spacetime metric, the Kalb-Ramond field and a dilaton. First one
extends the spacetime by introducing a set of winding coordinates $\tilde x_i$ dual
to the usual coordinates $x^i$. Here $i=1,\ldots ,D$ and is therefore a
$D$-dimensional $SL_D$ index. These coordinates are combined so that they fit into the
vector representation of $O_{D,D}$,
\beq
x^a = \begin{pmatrix} x^i \\ \tilde x_i \end{pmatrix}\,.
\eeq
Note that we use lowercase Latin indices $a,b,c,\dots$ from the start of the alphabet as $O_{D,D}$ vector indices, running from $1$ to $2D$.

One then finds that there is a generalised metric $M_{ab}$ \cite{Duff:1989tf} that can be written in terms of the $D$-dimensional fields $g_{ij}$ and $B_{ij}$.
Explicitly,
\beq
M_{ab}=\begin{pmatrix}  g_{ij} -  B_{ik} g^{kl} B_{lj} & B_{il}g^{ln} \\ -g^{mk}B_{kj} &  g^{mn} \end{pmatrix}\,, \label{eq:genmetric}
\eeq
where $g^{ij}$ is the inverse of the metric tensor $g_{ij}$.
We note that the inverse generalised
metric is then given by
\beq
M^{ab} = \begin{pmatrix} g^{ij} & -g^{ik}B_{kn} \\ B_{mk}g^{kj} & g_{mn} - B_{mk}g^{kl}B_{ln}\,.
\end{pmatrix}.\label{eq:geninvmetric} \eeq

At each point of the doubled space, $M_{ab}$ is an element of a symmetric space
with $G=O_{D,D}$ and $H=O_{D-1,1}\times O_{D-1,1}$ for the Lorentzian theory
or for the Euclidean version $H=O_D \times O_D$. A consequence of this is that
one can construct a vielbein for this doubled spacetime by writing
\beq
M_{ab}=e^\mu{}_a e^\nu{}_b M_{\mu\nu} \label{eq:genvielbein}\,,
\eeq
where $M_{\mu\nu}$ is a $2D \times 2D$ tangent space metric. For convenience
we will take $M_{\mu\nu}$ to be
\beq
M_{\mu\nu} = \begin{pmatrix} \bar\eta_{\alpha \beta} & 0 \\ 0 & \bar\eta^{\alpha \beta} \end{pmatrix}\,,
\eeq
where $\bar\eta_{\alpha \beta}$ is for the Lorentzian case the $D$-dimensional Minkowski spacetime metric and for
the Euclidean case the $D$-dimensional Kronecker delta. One sees from ~(\ref{eq:genvielbein})
that if a local $H$-transformation is made on $e^\mu{}_a$, the generalised metric
$M_{ab}$ is invariant.

Suppose that the vielbein of the $D$-dimensional metric is chosen to be $e^\alpha{}_i$.
Then a convenient choice of the $2D$-dimensional vielbein is
\beq e^\mu{}_a =
\begin{pmatrix}
 e^\alpha{}_i & 0 \\
 - e_{\alpha}{}^{j} B_{ji} & e_{\alpha}{}^{k}
\end{pmatrix}\,.
\label{eq:Vielbein}
\eeq
In making the choice of a lower triangular vielbein we have partially fixed the
local $H$-transformations. The corresponding inverse vielbein is therefore upper triangular
and given by
\beq e_\mu{}^a = \begin{pmatrix} e_\alpha{}^i & -e_\alpha{}^k B_{kj}\\ 0 & e^{\alpha}{}_{j}
\end{pmatrix} \,.\eeq

From here on, all generalised spacetime indices will be raised and lowered using the
generalised metric $M_{ab}$. All generalised tangent space indices will be raised and lowered
using the generalised tangent space metric $M_{\mu\nu}$.

The generalised metric $M_{ab}$ is an element of $O_{D,D}$ and thus our theory will also contain a quadratic form $\eta_{ab}$ that defines an $O_{D,D}$ structure.
(This new $\eta$ should not be confused with the ordinary Minkowski tangent space metric $\bar\eta$). A convenient representation of $\eta_{ab}$ is
\beq
\eta_{ab}= \begin{pmatrix} 0 & 1 \\ 1 & 0 \end{pmatrix} \label{eq:invariant} \,
\eeq
where the entries above are each $D \times D$-dimensional blocks.
The condition that $M_{ab} \in O_{D,D}$ then translates into
\beq
M^{ab}=\eta^{ac}M_{cd}\eta^{db}\,.
\eeq
One can easily verify that the generalised metric and its inverse, given by (\ref{eq:genmetric})
and (\ref{eq:geninvmetric}), obey this relation.

\section{Lie Derivatives, Tensors and Generalised Diffeomorphisms}
\label{sec:3}

In general relativity, the Lie derivative describes how a vector transforms under an infinitesimal diffeomorphism. Suppose that the diffeomorphism is generated by a
vector field $U^i$, then the Lie derivative $(L_UV)^i$ of a vector $V^i$ is given by the commutator of the vectors $U^i$ and $V^i$,
\beq
(L_UV)^i = [U, V]^i =  U^j\partial_j V^i - V^j\partial_j U^i\,.
\eeq
The first term in this expression is the result of the transportation of the vector field itself whereas the second term is a $GL_D$ transform on the components of the
vector $V^i$. As is well-known, the Lie derivative defines the algebra of diffeomorphisms: the commutator of two Lie derivatives is the Lie derivative of the commutator of the two vector fields,
\beq
[L_U,L_V] = L_{[U,V]}\,.
\eeq
One can extend the definition of the Lie derivative to arbitrary tensors by requiring that the Lie derivative of a scalar field $S$ just involves the transport term
\beq
L_U S= U^i\partial_i S \, ,
\eeq
and that the Lie derivative obeys the Leibniz rule. It should be noted that the Lie derivative maps tensors into tensors.

One can pursue a similar course in generalised geometry. Given two generalised vector $U^a$ and $V^a$, the generalised Lie derivative $(\mathcal{L}_U V)^a$ is defined as follows \cite{Hull:2009zb,Hull:2009mi,Berman:2012vc}
\beq
(\mathcal{L}_UV)^a = U^b \partial_b V^a - (V^b \partial_b U^a - Y^{ab}{}_{cd}V^d\partial_b
U^c)\,,
\label{eq:LieDerivative}
\eeq
where
\beq Y^{ab}{}_{cd} = \eta^{ab} \eta_{cd}\,. \eeq
Just like the ordinary Lie derivative the first term is a transport term, but now the second term is an $O_{D,D}$ transformation. It is important to note that the generalised Lie derivative preserves the $O_{D,D}$ structure
\begin{equation}
 \mathcal{L}_U \eta_{ab} = 0 \, .
\end{equation}
The generalised Lie derivative also defines an algebra of symmetries. Consider the commutator of two generalised Lie derivatives acting on a generalised vector $T^a$, then
\beq
\begin{split}
[\mathcal{L}_U,\mathcal{L}_V]T^a = \mathcal{L}_{\llbracket U,V \rrbracket} T^a  +  Y^{bc}{}_{de}\Big( & -(\partial_cV^d)(\partial_bU^a)T^e+(\partial_cU^d)(\partial_bV^a)T^e \\
& - {\frac{1}{2}}(\partial_bT^a)(\partial_cU^d) V^e
+ {\frac{1}{2}}(\partial_bT^a)(\partial_cV^d) U^e \Big)
\end{split}
\label{eq:LieClosure}
\eeq
where
\beq
\llbracket U,V \rrbracket^a = U^b\partial_b V^a - V^b\partial_b U^a +{
\frac{1}{2}}Y^{ab}{}_{cd}( V^d \partial_b
U^c - U^d \partial_b V^c )\,,
\eeq
is the Courant bracket of the two vectors $U^a$ and $V^a$ \cite{Hull:2009zb}.
The generalised Lie derivative therefore does not produce a closed algebra for arbitrary $U^a$ and $V^a$. To make the algebra closed we must restrict our theory. One way of achieving this is by imposing the section condition ~\cite{Hull:2009mi}
\beq
\eta^{ab} \partial_a \partial_b = 0\,.
\eeq
where the differential operator on the left can act on any of the fields and also on all products of fields. 

This condition can be interpreted as follows. The $2D$-dimensional generalised spacetime is not the physical spacetime. To find a physical spacetime we impose the section condition to bring us down to $D$ dimensions. Solving the section condition is equivalent to picking a global duality frame in the theory. Thus we end up matching the doubled theory to usual supergravity after solving the section condition. The approach taken here is to
maintain  manifest $O_{D,D}$ symmetry and thus keep the section condition as a constraint that must be solved. One obvious solution of the section condition is that
all fields are chosen to depend only on $x^i$ and not on $\tilde x_i$. In this case, everything must reduce back to the usual spacetime description based on
general relativity.

We shall say that two objects are \emph{weakly equal} if they are equal \emph{up to terms that vanish by the section condition}, and denote this by the symbol ``$\approx$''. If two things are equal without having to use the section condition we shall say they are \emph{strongly equal}, denoted by the usual equality symbol. This means we write $\eta^{ab}\partial_a\partial_bX\approx 0$ for any object $X$ in the theory. Since $X$ may be a product, $X=UV$, this also means that $\eta^{ab}\partial_aU \partial_bV \approx 0$ for any $U$ and $V$.

Returning to \eqref{eq:LieClosure} we then have
\beq [\mathcal{L}_U,\mathcal{L}_V] \approx \mathcal{L}_{\llbracket U,V \rrbracket}\,,
\eeq
and so the algebra of generalised diffeomorphisms closes up to terms that vanish
by the section condition.

Just as is done for the ordinary Lie derivative, one can define a generalised Lie derivative
acting on tensors of arbitrary type by asking for the Leibniz rule to be satisfied and
for the generalised Lie derivative of a scalar $S$ to just contain the transport term
\beq
\mathcal{L}_US = U^a\partial_a S\,.
\label{eq:LieScalar}
\eeq
As an example, the generalised Lie derivative of a
co-vector $W_a$ is
\beq (\mathcal{L}_UW)_a = U^b\partial_b W_a
+W_b \,\partial_a U^b - Y^{bc}{}_{da}W_b \,\partial_c U^d \,.\label{eq:LieDerivCo}
\eeq

This gives us an opportunity to ask how one recognises a tensor in generalised geometry.
The answer is simply that it is an object that transforms under an infinitesimal generalised
diffeomorphism like a generalised tensor. That is to say, that the infinitesimal
transformation must be that of the generalised Lie derivative.  Tensors may therefore
be either strongly tensorial or weakly tensorial, depending on whether the section condition
has been used in establishing the result.

As an example, consider a scalar $S$ transforming as in \eqref{eq:LieScalar}. In ordinary differential geometry, the partial derivative of a scalar then automatically transforms as a co-vector, that is $\delta_U \partial_a S = L_U \partial_a S$. However in the doubled theory, using the definition \eqref{eq:LieDerivCo}, we have
\begin{equation}
\delta_U \partial_a S - \mathcal{L}_U \partial_a S =
Y^{bc}{}_{ad} (\partial_c U^d) (\partial_b S) \,.
\end{equation}
This expression vanishes by the section condition and so $\delta_U \partial_a S \approx \mathcal{L}_U \partial_a S$. We then say that $\partial_a S$ is weakly a covector. In general, the generalised Lie derivative of a generalised tensor will only weakly be a generalised tensor.

%One might at this stage wonder about the role of the $O_{D.D}$ invariant $\eta$. It was
%defined in any coordinate system by ~(\ref{eq:invariant}). One might wonder if
%this was consistent with it being a generalised tensor. However, if one calculates its
%generalised Lie derivative, one finds that it is strongly zero

It is important to note at this point that there is an alternative to using the section condition in the theory. One may instead apply a generalised Scherk-Schwarz ansatz \cite{Scherk:1979zr} to the fields as described in \cite{Dall'Agata:2007sr,Aldazabal:2011nj,Geissbuhler:2011mx,Grana:2012rr,Berman:2012uy}. This manifestly does not satisfy the section condition but leads instead to an alternative set of constraints.  We will discuss this possibility in more detail in section \ref{sect:ScherkSchwarz}.

\section{The Action and the Connection}
\label{sec:4}

\subsection{The action of double field theory}

In the above we have described the global and local symmetries, the generalised metric
and the section condition, which allows all of this to work. In general relativity one would proceed from the metric
to construct a torsion-free, metric-compatible connection. Using that connection
one could then produce the Riemann curvature and use it to construct the
Ricci scalar for the action. The Ricci scalar is the only diffeomorphism invariant
object in Riemannian geometry that can be constructed only from the metric
with no more than two derivatives.

In DFT, one can find an action constructed only from the
generalised metric and doubled dilaton and their derivatives. It is invariant under generalised
Lie derivatives and the global $O_{D,D}$ duality group. The action is given by
\begin{equation}
 S = \int dx d\tilde{x} \,  e^{-2d} \, L \,,
\label{eq:Action}
\end{equation}
where the Lagrangian $L$ is
\begin{equation}
 \begin{split}
  L &= \frac{1}{8} M^{ab} \partial_a M^{cd} \partial_b M_{cd} - \frac{1}{2} M^{ab} \partial_a M^{cd} \partial_c M_{bd} +4 M^{ab} \partial_a \partial_b d - \partial_a \partial_b M^{ab} \\
  & \quad - 4 M^{ab} \partial_a d \partial_b d + 4 \partial_a M^{ab} \partial_b d
+ \frac{1}{2} \eta^{ab} \eta^{cd} \partial_a e^\mu{}_c \partial_b e_{\mu d}\,.
 \end{split}
\end{equation}
Here the doubled dilaton $d$ is related to usual dilaton $\Phi$ by
\beq d = \Phi - {\frac{1}{4}}\ln(|\det g_{ij}|)  \, ,\eeq
and is such that $e^{-2d}$ is a scalar of weight one in the generalised geometry. 

Suppose one solves the section condition by choosing all the fields to depend only on $x^i$ and makes a gauge choice for the vielbein as in \eqref{eq:Vielbein}. Then 
Then after carrying out the trivial $\tilde x_i$ integration, which produces an irrelevant volume factor, we are left with the standard effective action in string frame (\ref{eq:action}). 

One should note that $L$ is only \emph{weakly} an $O_{D,D}$ scalar. It is also only weakly $H$-invariant due to the final term. This term is not present in the original formulation of DFT \cite{Hohm:2010pp} but is required when considering a Scherk-Schwarz reduction in order to reproduce the correct gauged supergravity action \cite{Grana:2012rr}.

The action \eqref{eq:Action} is a great success in that its dynamics are manifestly $O_{D,D}$ invariant, it is weakly a generalised scalar and that it reproduces the correct low-energy action in the usual duality frame. It has, however, not been constructed geometrically: it is not manifestly made from the curvature or some generalised version of it.

There have been a number of different approaches to this problem, both for $O_{D,D}$ \cite{Jeon:2010rw,Jeon:2011cn,Hohm:2011si,Hohm:2012mf} and in the closely related $E_{D}$ case relevant to M-theory \cite{Coimbra:2011ky,Park:2013gaj, Cederwall:2013naa, Aldazabal:2013mya}, all of which are certainly legitimate and yet lead to connections with some undesirable properties.
In \cite{Jeon:2010rw,Jeon:2011cn,Park:2013gaj,Cederwall:2013naa} the approach is only partially covariant meaning the connection only gives a covariant derivative when acting on a subset of fields; in general this derivative is not covariant. In \cite{Hohm:2011si, Aldazabal:2013mya} the connection is not completely determined and as a result the generalised curvature tensor contains undetermined components. It can be contracted using a projection constructed using the $O_{D,D}$ structure to give the scalar that appears in the action but it is not a simple metric conntraction. Ideally we would like the formalism to only use the vielbein (and hence metric), derivatives and the so called Y-tensor in any formalism. If one restircts to using these objects then the structures generalise easily to the extended geometries of the excpetional groups. 

At this stage, let us step back and summarise the properties that a connection for double field theory could have: we might want it to
\begin{itemize}
\item define a covariant derivative that maps generalised tensors into generalised tensors,
\item be compatible with the generalised metric $M_{ab}$,
\item be compatible with the $O_{D,D}$ structure $\eta_{ab}$,
\item be completely determined in terms of the physical fields, in particular the vielbein and its derivatives,
\item be torsion-free,
\item lead to a curvature that may be contracted with the metric to give the scalar which appears in the action.
\end{itemize}

We will show in what follows how to find a connection that fulfills the first four of these conditions. It will however have vanishing curvature but non-vanishing torsion. All of the geometry is thus contained in the torsion. The problem of constructing the action from the curvature will then be replaced by that of constructing the action from the torsion. Furthermore, the connection will not be invariant under local $H$-transformations. We will show how to construct the action in terms of the torsion by demanding invariance under the local symmetry.

\subsection{The Weitzenb\"ock connection}

We define the covariant derivative in double field theory by the usual expression,
\begin{equation}
 \nabla_a V^b = \partial_a V^b + \Gamma^b_{ac} V^c \,,
\end{equation}
where $\Gamma^a_{bc}$ is the connection.
We want this object to define a tensor. Under a generalised diffeomorphism generated
by $U^a$,
\begin{equation} \label{eq:VarCovDev}
 \delta_U \nabla_a V^b = \mathcal{L}_U \partial_a V^b + Y^{cd}{}_{ae} \partial_d U^e \partial_c V^b + \left( Y^{bd}{}_{ce} \partial_a \partial_d U^e - \partial_a \partial_c U^b \right) V^c + \delta_U ( \Gamma^b_{ac} V^c )\,.
\end{equation}
Thus the covariant derivative transforms \emph{weakly} as a generalised tensor if
\begin{equation} \label{eq:VarConn}
 \delta_U \Gamma^a_{bc} \approx \mathcal{L}_U \Gamma^a_{bc} +  \partial_b \partial_c U^a - Y^{ad}{}_{ce} \partial_b \partial_d U^e \,.
\end{equation}
One may have wanted to define a connection that gives a \emph{strongly} covariant derivative.
This is not in general possible: \eqref{eq:VarCovDev} will only ever give a
weakly covariant transformation because the second term on the right-hand
side can only vanish when the section condition is used. As the connection will
only ever give a weakly covariant derivative, there is no reason to require the
connection to obey \eqref{eq:VarConn} strongly. (In section \ref{sect:ScherkSchwarz} we will consider the
case where the section condition is not obeyed but the Scherk-Schwarz reduction is used,
and then its consistency leads to other constraints \cite{Dall'Agata:2007sr,Aldazabal:2011nj,Geissbuhler:2011mx,Grana:2012rr,Berman:2012uy}.)

The key point of this paper is that so-called Weitzenb\"ock connection for the doubled geometry has many of the desired properties advertised in the previous section, and can be used to construct the action. 
The Weitzenb\"ock connection is given by
\begin{equation} \label{eq:Wbock}
 \Gamma^a_{bc} = e_\mu{}^a \partial_b e^\mu{}_c  \, .
\end{equation}
It obeys the transformation law \eqref{eq:VarConn} weakly, and therefore defines a suitable connection for double field theory. This connection is compatible with the generalised metric $M_{ab}$ and the $O_{D,D}$ structure $\eta_{ab}$. It has zero curvature but non-zero torsion. The absence of curvature may be viewed as a problem for describing the dynamics of the theory: naively, one would expect the action to be constructed from the curvature. In fact, as we show in the following section, the torsion of this connection alone is sufficient to construct the DFT action.

The Weitzenb\"ock connection was first discussed by Cartan in 1922 \cite{Cartan:torsion} and later by
Einstein in 1928 \cite{Einstein:torsion} in attempts at
reformulating general relativity in a  way useful for the unification of gravity with
electromagnetism. The Weitzenb\"ock connection also appears naturally
when one carries out the nonlinear
realisation of $G/H$. Its appearance in the theory of nonlinear realisations
is discussed in \cite{Berman:2011jh} amongst other places. Given that one may construct
the generalised vielbein and hence generalised metric using the theory of nonlinear
realisations it seems natural to ask about the role of this resulting connection, which in the language of \cite{Berman:2011jh} is
\beq
\Gamma= g_E^{-1}d g_E \, .
\eeq

Let us now briefly verify the important properties of the Weitzenb\"ock connection.
Recalling the definition of the
generalised metric in terms of the vielbein, one can easily show that it is a metric
connection,
\begin{equation}
 \nabla_a M_{bc} = 0\,.
\end{equation}
We can also ask for the covariant derivative of the vielbein to vanish. This is used to define
the spin connection $\omega_a{}^{\mu\nu}$.
The covariant derivative of the vielbein is, just like in general relativity, defined by
\beq
\nabla_a e^\mu{}_b = \partial_a e^\mu{}_b - \Gamma^c_{ab}e^\mu{}_c - \omega_a{}^\mu{}_\nu e^\nu{}_b = 0\,.
\eeq
We now see that substituting in the Weitzenb\"ock connection yields a vanishing spin
connection
\begin{equation}
 \omega_a{}^{\mu\nu} = 0 \, .
\end{equation}

Finally, one can show that the Weitzenb\"ock connection is compatible with the $O_{D,D}$
structure $\eta_{ab}$. To do so we recall that the generalised metric is constrained to
obey $M^{ab} \eta_{bc} = \eta^{ab} M_{bc}$, which leads to the following constraint on
the vielbein
\begin{equation}
e_\mu{}^a = \eta^{ab} \eta_{\mu \nu} e^\nu{}_b \,,
\end{equation}
where $\eta_{\mu \nu} = e_\mu{}^a e_\nu{}^b \eta_{ab}$. 
The covariant derivative of the $O_{D,D}$ structure is
\begin{equation}
\begin{split}
\nabla_a \eta_{bc} & = - \Gamma^d_{ab} \eta_{dc} - \Gamma^d_{ac} \eta_{db} \\
 %& = - e_\mu{}^d \partial_a e^\mu{}_b \eta_{d c} - e_\mu{}^d \partial_a e^\mu{}_c \eta_{db} \\
 &  = 0 \,,
\end{split} 
\label{eq:NablaEta}
\end{equation}
using the above vielbein identity.

Note that any connection which preserves $\eta_{ab}$ satisfies
\begin{equation} \label{eq:YGamma}
 Y^{ad}{}_{ce} \Gamma^e_{bd} = - \Gamma^a_{bc} \,.
\end{equation}
This is equivalent to projecting the $(^a_{\,c})$ indices of $\Gamma^a_{bc}$ into the adjoint representation of $O_{D,D}$.

\subsection{Curvature and Torsion}

In general relativity, the Riemann curvature tensor of a connection is defined by
\begin{equation}
 R^a{}_{bcd} = \partial_c \Gamma^a_{db} - \partial_d \Gamma^a_{cb} + \Gamma^a_{ce} \Gamma^e_{db} - \Gamma^a_{de} \Gamma^e_{cb} \,. \label{eq:riemann}
\end{equation}
In generalised geometry, where the transformation property of the connection is (\ref{eq:VarConn}), this object is not strongly a tensor, having an anomalous transformation \cite{Hohm:2011si}
\beq
\Delta_U R^a{}_{bcd} = 2 Y^{ef}{}_{g[c} \partial_{d]} \partial_e U^g \Gamma^a_{fb} \,.
\label{eq:DeltaRiemann}
\eeq
Instead one is led to define a generalised curvature tensor
\begin{equation}
 \mathcal{R}^a{}_{bcd} = R^a{}_{bcd} + Y^{ae}{}_{df} R^f{}_{cbe} + Y^{ef}{}_{dg} \Gamma^g_{ec} \Gamma^a_{fb} \, . \label{eq:sriemann}
\end{equation}
This is strongly a tensor for a connection which obeys the strong version of its transformation law.

For the Weitzenb\"ock connection, \eqref{eq:riemann} is weakly a tensor and so we will refer to it as the Riemann curvature tensor, although in general this is a misnomer. In fact for the Weitzenb\"ock connection this Riemann tensor vanishes \emph{strongly}. We also find that the generalised curvature tensor \eqref{eq:sriemann} vanishes \emph{weakly}
\begin{equation}
 \mathcal{R}^a{}_{bcd} \approx 0 \, .
\end{equation}
We conclude that in generalised geometry the Weitzenb\"ock connection is flat, and so we cannot describe the dynamics through the curvature.

The torsion of a connection, defined by
\begin{equation}
 T_{bc}{}^a = \Gamma^a_{bc} - \Gamma^a_{cb}
\end{equation}
is not a generalised tensor. There is however a {\it{generalised torsion}} \cite{Coimbra:2011nw}, denoted $\tau_{bc}{}^a$, defined by
\begin{equation}
 \tau_{bc}{}^a U^b V^c
\equiv \left( \mathcal{L}^\nabla_U - \mathcal{L}^\partial_U \right) V^a\,,
\end{equation}
where $\mathcal{L}^\partial_U$ is the usual Lie derivative as defined in \eqref{eq:LieDerivative} and $\mathcal{L}^\nabla_U$ is the Lie derivative with all partial derivatives replaced by covariant ones. This gives
\begin{equation}
 \tau_{bc}{}^a = T_{bc}{}^a + Y^{ad}{}_{ce} \Gamma^e_{db} \label{eq:tau}
\end{equation}
as the generalised torsion of a connection $\Gamma^a_{bc}$. It is \emph{strongly} a generalised tensor if the connection obeys (\ref{eq:VarConn}) strongly. However, because the Weitzenb\"ock connection behaves only weakly as a connection, its generalised torsion is only weakly a tensor. For any connection preserving $\eta_{ab}$, the generalised torsion will be antisymmetric in its lower indices, $\tau_{bc}{}^a = \tau_{[bc]}{}^a$, which follows as a consequence of \eqref{eq:NablaEta}. It is this generalised torsion that we now hope to be able to use to obtain the dynamics.

We will demonstrate how this actually works below. For now let us explore the relationship between generalised torsion and curvature by studying the Ricci and Bianchi identities. 

The Ricci identity is obtained by looking at the commutator of two covariant derivative operators. For a generalised scalar $S$ this gives
\begin{equation}
 \nabla_a \nabla_b S - \nabla_b \nabla_a S = -T_{ab}{}^c \nabla_c S\,,
\label{eq:ScalarRicciId}
\end{equation}
while for a covector we have
\beq
(\nabla_a \nabla_b - \nabla_b \nabla_a) V_c = R^d{}_{cba}V_d - T_{ab}{}^d\nabla_d V_c\,,
\label{eq:CovRicciId}
\eeq
where the Riemann tensor is defined as in ~(\ref{eq:riemann}). Similar identities hold for vectors and tensors of higher rank.

The above Ricci identities are valid in generalised geometry with the caveat that the objects appearing on the right-hand side - the torsion and Riemann curvature - are not generalised tensors. On the left-hand side the commutator of two covariant derivatives is indeed a tensor. This suggests there should be some way of rewriting these identities in terms of generalised tensors.

For instance, consider the scalar Ricci identity, \eqref{eq:ScalarRicciId}. The right-hand side of this equation can be re-written as
\beq
-T_{ab}{}^c \nabla_c S = -\tau_{ab}{}^c \nabla_c S + Y^{cd}{}_{be}\Gamma^e_{da}\nabla_cS
\eeq
For the Weitzenb\"ock connection, the second term on the right hand side vanishes by
the section condition. Thus, for scalars, the Ricci identity reads
\beq
\nabla_a \nabla_b S - \nabla_b \nabla_a S \approx -\tau_{ab}{}^c \nabla_c S\,.
\eeq
Similarly, consider the Ricci identity for covectors \eqref{eq:CovRicciId}.
For the Weitzenb\"ock connection
the Riemann tensor vanishes and after using the section condition the torsion can be replaced
by the generalised torsion. Thus
\beq
(\nabla_a \nabla_b - \nabla_b \nabla_a) V_c \approx -\tau_{ab}{}^d \nabla_d V_c.
\eeq
Similar identities hold for tensors of any type.

Finally, we have the following Bianchi identity \cite{Hohm:2012mf}
\beq
3 \mathcal{R}^a{}_{[bcd]} = 3 \nabla_{[b} \tau_{cd]}{}^a - 3 \tau_{[bc}{}^e \tau_{d]e}{}^a - Y^{ae}{}_{bf} \nabla_e \tau_{cd}{}^f \, .
\eeq
For the Weitzenb\"ock connection, $\mathcal{R}^a{}_{bcd} \approx 0$, so this becomes
\beq
3 \nabla_{[b} \tau_{cd]}{}^a \approx 3 \tau_{[bc}{}^e \tau_{d]e}{}^a + Y^{ae}{}_{bf} \nabla_e \tau_{cd}{}^f \,.
\eeq

\subsection{Constructing the action}

We know that DFT should be invariant under both global $O_{D,D}$ and local $H \equiv O_D \times O_D$. Infinitesimally, such an $H$-transformation generated by $\lambda^{\mu}_{\ph{\mu}\nu}$ must satisfy
\begin{equation}
 \begin{split}
  \lambda^\mu_{\ph{\mu}\rho} M^{\rho \nu} + M^{\mu \rho} \lambda^\nu_{\ph{\nu}\rho} & \equiv \lambda^{\mu \nu} + \lambda^{\nu \mu} = 0\,, \\
  \lambda^\mu_{\ph{\mu}\rho} \eta^{\rho \nu} - \eta^{\mu \rho} \lambda_\rho^{\ph{\rho}\nu} &= 0\,,
 \end{split}
\end{equation}
where we raise tangent space indices with $M_{\mu \nu}$.

The Weitzenb\"ock connection \eqref{eq:Wbock} is not invariant under these local $H$-transformations
\begin{equation}
 \Delta_\lambda \Gamma^a_{bc} = e_{\mu}{}^a e_{\nu \, c} \partial_b \lambda^{\mu\nu}\,,
\end{equation}
and neither is the generalised torsion. Even though the generalised torsion is not invariant we may seek combinations of terms quadratic in the torsion that are locally $H$-invariant. Then we may construct a Lagrangian $L$ from these terms which is invariant under the local $H$ symmetry and is an $O_{D,D}$ scalar. Our DFT action will then be
\begin{equation}
 S = \int dx d\tilde{x} \,  e^{-2d} \, L\,.
\end{equation}

%It is easy to construct $O_{D,D}$ scalars involving two derivatives from the Weitzenb\"ock connection: %we just contract squares of the generalised torsion.

There are only four independent combinations quadratic in the torsion that one can consider because of the identity \eqref{eq:YGamma}
\begin{equation}
 \begin{split}
  \tau_{bc}{}^a \tau_{da}{}^c M^{bd} &= \Gamma^a_{bc} \Gamma^c_{da} M^{bd} - 4 \Gamma^a_{cb} \Gamma^c_{da} M^{bd} + 2 \Gamma^a_{cb} \Gamma^c_{ad} M^{bd} - 2 \eta^{ae} \eta_{cf} \Gamma^f_{eb} \Gamma^c_{ad} M^{bd}, \\
  \tau_{bc}{}^a \tau_{ef}{}^d M_{ad} M^{be} M^{cf} &= M_{ad} M^{be} M^{cf} \left( 3 \Gamma^a_{bc} \Gamma^d_{ef} - 6 \Gamma^a_{bc} \Gamma^d_{fe} \right), \\
  \tau_{bc}{}^a \tau_{ad}{}^c \eta^{bd} &= \eta^{bd} \left( 6 \Gamma^a_{bc} \Gamma^c_{ad} - 3 \Gamma^a_{bc} \Gamma^c_{da} \right), \\
  \tau_{bc}{}^a \tau_{ef}{}^d \eta_{ad} M^{be} M^{cf} &= 2 \eta_{ad} M^{be} M^{cf} \Gamma^a_{bc} \Gamma^d_{ef} - 2 M^{be} M^{cf} \eta_{ad} \Gamma^a_{bc} \Gamma^d_{fe} - 4 \eta^{cf} M^{be} M_{ad} \Gamma^a_{bc} \Gamma^d_{fe} \\
  & \quad + M_{ad} \eta^{be} M^{cf} \Gamma^a_{bc} \Gamma^d_{ef}\,.
 \end{split}
\end{equation}
For a strong connection, these would be strong $O_{D,D}$ scalars but for the Weitzenb\"ock connection, these are weak scalars.

In addition one can construct a weak $O_{D,D}$ covector by taking the covariant derivative of the dilaton $d$ (see appendix \ref{Weighted})
\begin{equation}
 \nabla_a d = \partial_a d + \frac{1}{2} \Gamma^c_{ca}\,.
\label{eq:covd}
\end{equation}
Using this we can also construct the following scalars
\begin{equation}
 \begin{split}
 M^{ab} \nabla_a d \nabla_b d &= \frac{1}{4} \Gamma^c_{ca} \Gamma^d_{db} M^{ad} + M^{ab} \Gamma^c_{ca} \partial_b d + M^{ab} \partial_a d \partial_b d, \\
  \eta^{ab} \nabla_a d \nabla_b d &= \frac{1}{4} \Gamma^c_{ca} \Gamma^d_{db} \eta^{ab} + \eta^{ab} \Gamma^c_{ca} \partial_b d + \eta^{ab} \partial_a d \partial_b d, \\
  M^{ab} \nabla_a \nabla_b d &= M^{ab} \partial_a \partial_b d + \frac{1}{2} M^{ab} \partial_a \Gamma^c_{cb} - M^{ab} \Gamma^c_{ab} \partial_c d - \frac{1}{2} M^{ab} \Gamma^c_{ab} \Gamma^d_{dc}, \\
  \eta^{ab} \nabla_a \nabla_b d &= \eta^{ab} \partial_a \partial_b d + \frac{1}{2} \eta^{ab} \partial_a \Gamma^c_{cb} - \eta^{ab} \Gamma^c_{ab} \partial_c d - \frac{1}{2} \eta^{ab} \Gamma^c_{ab} \Gamma^d_{dc}\,.
 \end{split}
\end{equation}
By calculating the transformation of these $O_{D,D}$ scalars under the $H$ symmetry, we find that the \emph{only} $H$-invariant combination is
\begin{equation} \label{eq:Lagrangian}
 L = - \frac{1}{12} \tau_{bc}{}^a \tau_{ef}{}^d M_{ad} M^{be} M^{cf} - \frac{1}{4} \tau_{bc}{}^a \tau_{da}{}^c M^{bd} - 4 M^{ab} \nabla_a d \nabla_b d + 4 M^{ab} \nabla_a \nabla_b d \,.
\end{equation}
The overall normalisation has been chosen to make contact with the standard literature. 

Inserting the expressions for the torsion and covariant derivative of $d$ we find that we can write \eqref{eq:Lagrangian} as
\begin{equation}
 \begin{split}
  L &= \frac{1}{8} M^{ab} \partial_a M^{cd} \partial_b M_{cd} - \frac{1}{2} M^{ab} \partial_a M^{cd} \partial_c M_{bd} +4 M^{ab} \partial_a \partial_b d - \partial_a \partial_b M^{ab} \\
  & \quad - 4 M^{ab} \partial_a d \partial_b d + 4 \partial_a M^{ab} \partial_b d + \frac{1}{2} \eta^{ab} \eta^{cd} \partial_a e^\mu{}_c \partial_b e_{\mu d}  \, .
 \end{split}
\end{equation}
This agrees with the original formulation \cite{Hohm:2010pp} up to the final term which vanishes by the section condition. This term is only \emph{weakly} $H$-invariant but is required to match the Scherk-Schwarz reduced theory with the gauged supergravity potential as discussed in \cite{Grana:2012rr}. There it had to be inserted by hand, but here it necessarily appears here as a consequence of our construction.

\section{Scherk-Schwarz and Weitzenb\"ock} \label{sect:ScherkSchwarz}
\label{sec:5}

The above discussion of double field theory depended crucially on the section condition for everything to make sense. This condition is a very restrictive one and it is natural to look for ways to weaken it. One such way is to use a \emph{Scherk-Schwarz reduction} \cite{Scherk:1979zr} on the DFT \cite{Dall'Agata:2007sr,Aldazabal:2011nj,Geissbuhler:2011mx,Grana:2012rr,Berman:2012uy,Dall'Agata:2008qz,Hull:2009sg}. In this section we will describe how our formalism based on the Weitzenb\"ock connection carries over to this viewpoint.

\subsection{Scherk-Schwarz reduction}

We begin with some double field theory which we refer to as the parent DFT. We perform a Scherk-Schwarz reduction \cite{Scherk:1979zr} by splitting the coordinates of the theory into ``internal'' coordinates, denoted collectively by $\mathbb{Y}$, and ``external'' coordinates denoted collectively by $\mathbb{X}$. We then reduce by demanding that the coordinate dependence of all the fields in the parent DFT factorise in a particular way. Namely, we assume that tensors $V^a{}_b(\mathbb{X}, \mathbb{Y})$ and the dilaton $d(\mathbb{X}, \mathbb{Y})$ obey
\beq
V^a{}_b ( \mathbb{X}, \mathbb{Y} ) = ( W^{-1} )^a{}_A ( \mathbb{Y} )  W^B{}_b (\mathbb{Y}) \hat{V}^A{}_B (\mathbb{X} )
\quad , \quad
d(\mathbb{X}, \mathbb{Y} ) = \hat{d} (\mathbb{X} ) + \lambda ( \mathbb{Y} ) \,,
\label{SSreds}
\eeq
where $W^A{}_a \in O_{D,D}$.
In general we will leave the dependence on coordinates implicit, and denote any object which depends only on $\mathbb{X}$ with a hat.
% It is these objects that arise in the gauged double field theory we are interested in studying.
Capital latin indices $A,B,C \dots$ are used for this gauged DFT, while the indices $a,b,c, \dots$ to be associated with the parent DFT on which we are applying the Scherk-Schwarz reduction.

The matrix $W^A{}_a$ and scalar twist $\lambda$ which specify the Scherk-Schwarz reduction enter into the gauged double field theory only in the particular combinations
\beq
f_{BC}{}^A = 3 Y^{AE}{}_{D[E} (W^{-1})^a{}_B (W^{-1})^b{}_{C]} \partial_a W^D{}_b
\quad , \quad
f_A = \partial_a (W^{-1})^a{}_A - 2 ( W^{-1} )^a{}_A \partial_a \lambda  \,.
\label{SSgaugings}
\eeq
These are known as Scherk-Schwarz gaugings. Since we do not want any dependence on the internal coordinates $\mathbb{Y}$ to enter the reduced theory, we must take the gaugings to be constant. Note that $f_{BC}{}^A$ can be written as $f_{BC}{}^A = 3 \Omega_{[DB}{}^E Y^{AD}{}_{C]E}$ where
\beq
\Omega_{AB}{}^C = (W^{-1})^a{}_A (W^{-1})^b{}_B \partial_a W^C{}_b = - \Omega_{AE}{}^D Y^{CE}{}_{DB} \,.
\eeq
Consistency of the theory further requires that acting on any hatted quantity $\hat{g}(\mathbb{X})$ we have \cite{Grana:2012rr}
\beq
\partial_b \, \hat{g}(\mathbb{X}) = W^A{}_b \partial_A \hat{g} (\mathbb{X} )
\label{eq:SSAssumedConstraints}
\eeq
and we assume, again following \cite{Grana:2012rr}, that the duals of external and internal coordinates are respectively also external and internal. This can be realised by
\beq
\eta^{ab} \partial_a W^A{}_c \partial_b \hat{g}( \mathbb{X} ) = 0 \,.
\label{eq:SSAssumedConstraints1}
\eeq
On the gaugings these constraints lead to
\beq
 f_{BC}{}^A \partial_A \hat{g}(\mathbb{X}) = 0
\quad , \quad \eta^{AB} f_A \partial_B \hat{g} ( \mathbb{X} ) = 0 \,.
\label{eq:AssumedGaugingsConstraints}
\eeq
The Lie derivative of a tensor in the parent DFT will induce a gauge transformation in the gauged DFT \cite{Grana:2012rr} given by
\begin{equation}
 \mathcal{L}_U V^a{}_b = W^B{}_b \WI^a{}_A \hat{\mathcal{L}}_{\hat{U}} \hat{V}^A{}_B \, , \label{eq:GaugedLie}
\end{equation}
where
\begin{equation}
 \begin{split}
  \hat{\mathcal{L}}_{\hat{U}} \hat{V}^A{}_B &\equiv \hat{U}^C \partial_C \hat{V}^A{}_B + \left( Y^{AD}{}_{CE} \partial_D \hat{U}^E - \partial_C \hat{U}^A \right) \hat{V}^C{}_B + \left( \partial_B \hat{U}^C - Y^{CD}{}_{BE} \partial_D \hat{U}^E \right) \hat{V}^A{}_C \\
  & \quad - f_{CD}{}^A \hat{U}^C \hat{V}^D{}_B + f_{CB}{}^D \hat{U}^C \hat{V}^A{}_D \, .
 \end{split}
\label{eq:GaugedLieHat}
\end{equation}
Requiring closure of the gauge algebra generated by this Lie derivative leads to further constraints on our theory \cite{Grana:2012rr}. In particular, we find we should impose the section condition in the external space
\beq
\eta^{AB} \partial_A \hat{g}(\mathbb{X}) \partial_B \hat{h}(\mathbb{X})  = 0 \,,
\label{eq:SSGaugedSecCond}
\eeq
as well as Jacobi identities on the gaugings
\beq
 f_{[AB}{}^E f_{C]D}{}^F = 0 \quad , \quad
 \Omega_{E[A}{}^G \Omega_{|F|B}{}^D Y^{EF}{}_{C]G} = 0 \, .
\label{eq:SSJacobi}
\eeq

\subsection{Reduction of the connection} \label{sect:SSreduct}

Let us now examine how the Weitzenb\"ock connection behaves under this reduction. The Scherk-Schwarz twist for the vielbein is
\beq
e^\mu{}_a  = W^B{}_a  \hat{e}^\mu{}_B
\quad , \quad
e_\mu{}^a = (W^{-1})^a{}_B \hat{e}_\mu{}^B
\eeq
leading to
\beq
%\begin{split}
\Gamma^a_{bc} %e_\mu{}^a \partial_b e^\mu{}_c & = (W^{-1})^a_A W^B_b W^C_c \hat{e}_\mu{}^A \partial_B \hat{e}^\mu{}_C  + (W^{-1})^a_A \partial_b W^A_c \\
% & = (W^{-1})^a_A W^B_b W^C_c \left( \hat{\Gamma}^A_{BC} + (W^{-1})^d_B (W^{-1})^e_C \partial_d W^A_e \right) \\
  = (W^{-1})^a{}_A W^B{}_b W^C{}_c \left( \bar{\Gamma}^A_{BC} + \Omega_{BC}{}^A \right),
%\end{split}
\eeq
where we have defined $\bar{\Gamma}^A_{BC} (\mathbb{X}) = \hat{e}_\mu{}^A \partial_B \hat{e}^\mu{}_C$, which is the Weitzenb\"ock connection associated with the hatted vielbein.
However, we also have
\beq
%\begin{split}
\partial_b V^a % & = ( \partial_b (W^{-1})_a^A) \hat{V}^A + (W^{-1})^a_A W^B_b \partial_B \hat{V}^A \\
% &  = (W^{-1})^a_A W^B_b \left( \partial_B \hat{V}^A +  (W^{-1})^d_B \partial_d (W^{-1})^R_C W^A_R \hat{V}^C  \right) \\
% & = (W^{-1})^a_A W^B_b \left( \partial_B \hat{V}^A - (W^{-1})^d_B (W^{-1})^R_C \partial_d W^A_R \hat{V}^C \right) \\
% &
 = (W^{-1})^a{}_A W^B{}_b \left( \partial_B \hat{V}^A - \Omega_{BC}{}^A \hat{V}^C \right)\,,
%\end{split}
\eeq
so that
\beq
\nabla_a V^b = (W^{-1})^b{}_A W^B{}_a \hat{\nabla}_B \hat{V}^A
\eeq
with
\beq
\hat{\nabla}_B \hat{V}^A =  \partial_B \hat{V}^A + \bar{\Gamma}^A_{BC} \hat{V}^C \,.
\eeq
It can be checked that this object transforms covariantly under the diffeomorphisms in the gauged theory generated by \eqref{eq:GaugedLieHat}. Hence the Weitzenb\"ock connection also defines a covariant derivative in the gauged DFT.

The generalised torsion becomes
\beq
\begin{split}
\tau_{bc}{}^a & %=  (W^{-1})^a_A W^B_b W^C_c \left( \hat{\tau}_{BC}{}^A + \Omega_{BC}{}^A - \Omega_{CB}{}^A + Y^{AF}{}_{CE} \Omega_{FB}{}^E \right) \\
% &
  \equiv (W^{-1})^a{}_A W^B{}_b W^C{}_c \hat{\tau}_{BC}{}^A \\
 & =  (W^{-1})^a{}_A W^B{}_b W^C{}_c \left( \bar{\tau}_{BC}{}^A + f_{BC}{}^A \right) \,,
\end{split}
\eeq
where we have again defined $\bar{\tau}_{BC}{}^A$ using the Weitzenb\"ock connection $\bar{\Gamma}_{BC}^A$. It is noteworthy that the reduced generalised torsion $\bar{\tau}_{BC}{}^A$ and the structure constants $f_{BC}{}^A$ appear on the same footing.

One might ask whether $\bar{\tau}^A_{BC}$ alone is a tensor in the gauged DFT. The induced transformation for $\hat{\tau}^A_{BC}$ is given by the induced Lie derivative
\begin{equation}
 \hat{\delta}_{\hat{U}} \hat{\tau}_{BC}{}^A = \hat{\mathcal{L}}_{\hat{U}} \hat{\tau}_{BC}{}^A \, .
\end{equation}
Because $f_{BC}{}^A$ are constants, their variation has to vanish and thus
\begin{equation}
 \hat{\delta}_{\hat{U}} \bar{\tau}_{BC}{}^A = \hat{\mathcal{L}}_{\hat{U}} \bar{\tau}_{BC}{}^A \, ,
\end{equation}
and $\bar{\tau}_{BC}{}^A$ is also a tensor in the gauged DFT.

Finally, we should investigate the reduction of the covariant derivative of the dilaton, equation \eqref{eq:covd}. In the Scherk-Schwarz reduction,
\beq
\nabla_a d = W^A{}_a \left( \hat{\nabla}_A \hat{d}  - \frac{1}{2} f_A \right) \,,
\eeq
where $f_A$ is the gauging of the dilaton, defined in \eqref{SSgaugings}, and we have defined
\beq
\hat{\nabla}_A \hat{d} = \partial_A \hat{d} + \frac{1}{2} \bar{\Gamma}^B_{BA} \, .
\label{eq:NablaHatdHat}
\eeq
Note that although we have written this quantity as a covariant derivative it will not in fact transform covariantly in the gauged DFT unless $f_A = 0$ (see appendix \ref{Weighted}). Fortunately, this is one of the conditions necessary to obtain a gauge invariant action.

% move to appendix?
%It can be checked that this leads to
%\beq
%\nabla_a \nabla_b d = W_b^C W_a^B \hat{\nabla}_B \hat{\nabla}_C \hat{d}  + \frac{1}{2} W_a^B W_b^C \hat{\Gamma}^A_{BC} f_A - \frac{1}{2} W_b^C \partial_a f_C
%\eeq
%The last term vanishes as $f_A$ is taken to be constant. We will use these results below to construct the action for gauged double field theory.

\subsection{Discussion of weak tensorial properties}

When we carry out a Scherk-Schwarz reduction we replace the section condition of the parent DFT with a weaker set of constraints acting on the twist variables $W^A{}_a$, $\lambda$ and on the gauged DFT variables. One might have hoped that all the objects we used to construct the parent DFT action were strong tensors. Then the Scherk-Schwarz ansatz would clearly be consistent. However, we are not so lucky: we were only able to construct weak tensors. This means that the Scherk-Schwarz ansatz, where the section condition is relaxed, may not be valid. In \cite{Grana:2012rr}, it was shown explicitly that the generalised Scherk-Schwarz ansatz does work - that is to say, produces a gauged DFT action which is indeed a scalar. We are going to show how this follows from our formalism. 

The complete set of Scherk-Schwarz constraints following from \cite{Grana:2012rr} is
\begin{equation}
\label{eq:SSConstraints}
\begin{split}
% \partial_b \hat{g}(\mathbb{X}) & = W^A{}_b \partial_A \hat{g} (\mathbb{X} ) \,,\\
 \eta^{ab} \partial_a W^A{}_c \partial_b \hat{g}(\mathbb{X}) & \ssapprox 0 \, ,\\%  \left( W^{-1} \right)^a_{A} \partial_a \hat{g}(\mathbb{X}) &= \partial_A \hat{g}(\mathbb{X}), \\
 %f^A{}_{BC} \partial_A \hat{g}( \mathbb{X} ) & = 0 = f^A \partial_A \hat{g}(X) \,, \\
 \eta^{AB} \partial_A \hat{g}(\mathbb{X}) \partial_B \hat{h}(\mathbb{X}) &\ssapprox 0 \, , \\
 f_{[AB}{}^E f_{C]D}{}^F & \ssapprox 0  \, , \\
 \Omega_{E[A}{}^G \Omega_{|F|B}{}^D Y^{EF}{}_{C]G} &\ssapprox  0 \, , \\
 f_A & \ssapprox  0 \,.
\end{split}
\end{equation}
where $\hat{g}(\mathbb{X}), \hat{h}(\mathbb{X})$ are any fields which only depend on $\mathbb{X}$. We will also need to impose constancy of $\Omega_{AB}{}^C$ in order for the covariant derivative of the dilaton to remain a weak tensor. Note that this is stronger than requiring $f_{AB}{}^C$ to be constant. In fact the constancy of $\Omega_{AB}{}^C$  has been observed and analysed before in the context of generalised Scherk-Schwarz reductions \cite{Berman:2012uy}. 

We take this set of constraints to be the replacement of the section condition for the Scherk-Schwarz reduced theory. We shall use the symbol $\ssapprox$ to denote \emph{equality up to the Scherk-Schwarz constraints} (``SS-weakly equal'').

Let us now analyse our weak tensors from the Scherk-Schwarz perspective and show that they transform as tensors when the Scherk-Schwarz constraints are used. This will show that the parent DFT action will define an $O_{D,D}$ scalar even if the section condition is relaxed as in the Scherk-Schwarz theory. We will first carry out the variations in the parent DFT, and then impose the Scherk-Schwarz ansatz for the gauge parameters $U^a$ and the other tensors appearing:
\beq
U^a = (W^{-1})^a{}_A \hat{U}^A \, .
\eeq
For the covariant derivative of a vector the transformation $\delta_U \nabla_a V^b$ differs from the Lie derivative $\mathcal{L}_U \nabla_a V^b$ by two terms which in the parent DFT vanish upon the section condition. One of these terms arises from the partial derivative term in the covariant derivative, and the other from the partial derivative in the Weitzenb\"ock connection. This discrepancy can be written as
\begin{equation}
 \Delta^{\eta}_U \nabla_a V^b \equiv \delta_U \nabla_a V^b - \mathcal{L}_U \nabla_a V^b = Y^{de}{}_{af} V^c  \partial_d U^f e_\mu{}^b \partial_e e^\mu{}_c + Y^{de}{}_{af} \partial_d U^f \partial_e V^b  \,,
\end{equation}
which is obviously $\approx 0$ in the parent DFT. If one now substitutes in the Scherk-Schwarz ansatz and then uses the constraints \eqref{eq:SSConstraints} we do indeed find that
\beq
\Delta^{\eta}_U \nabla_a V^b \ssapprox 0 \,,
\eeq
so the covariant derivative is still tensorial. The crucial feature of this verification is that the variation of the Weitzenb\"ock connection contains anomalous terms involving the parent section condition which by themselves do not vanish SS-weakly. They will only vanish SS-weakly when used with those anomalous terms from the partial derivative term.

This means we still need to check whether the generalised torsion is SS-weakly tensorial. Its anomalous terms are given by
\begin{equation}
\begin{split} 
  \Delta^{\eta}_U \tau_{bc}{}^a & = Y^{de}{}_{bf} e_\mu{}^a \partial_d U^f \partial_e e^\mu{}_c - Y^{de}{}_{cf} e_{\mu}{}^a \partial_d U^f \partial_e e^\mu{}_b + Y^{de}{}_{cf} e_\mu{}^f \partial_d U^a \partial_e e^\mu{}_b \\
 & \ssapprox 0 
\end{split}
\end{equation}
and can also be shown to SS-weakly vanish by the constraints \eqref{eq:SSConstraints}. Thus, the generalised torsion is SS-weakly a tensor and can be used to build $O_{D,D}$ scalars even when using the Scherk-Schwarz constraints rather than the section condition on the parent DFT.

The final object to discuss is the covariant derivative of the dilaton, which after some use of the constraints can be shown to obey
\begin{equation}
  \Delta^{\eta}_U \nabla_a d \ssapprox  - \frac{1}{2} Y^{AE}{}_{CD} \WI^e{}_E W^B{}_a \hat{U}^D \partial_e \Omega_{AB}{}^C
\end{equation}
This vanishes if we require $\Omega_{AB}{}^C$ be constant. 

It is now clear that all objects appearing in the parent DFT action \eqref{parentDFT} are SS-weak tensors and thus their contraction will define an SS-weak scalar. The action can therefore be used for a Scherk-Schwarz ansatz, as has been checked in \cite{Grana:2012rr} by an explicit computation.

\subsection{The action}

We can now verify that carrying out the Scherk-Schwarz procedure on the parent DFT with our uniquely $H$-invariant Lagrangian \eqref{eq:Lagrangian} leads to the gauged DFT action previously constructed in \cite{Grana:2012rr}. If one were to start with the usual DFT action, then to make the Scherk-Schwarz reduction work, an additional term %Note that while this is somewhat unsurprising, as our action is of course the same as the usual DFT action, if one starts with the latter then to make the Scherk-Schwarz reduction work an additional term
\beq
\frac{1}{2} \eta^{ab} \partial_a e^\mu{}_c \partial_b e_{\mu \, d} \eta^{cd}
\label{dEdE}
\eeq
must be added to the Lagrangian by hand as was done in \cite{Grana:2012rr}. This term automatically appears in the Lagrangian \eqref{eq:Lagrangian}.

Let us write our parent DFT action as
\beq
S = \int dx d \tilde x  \, e^{-2d} \, L(\tau(e), e, d)\,,
\eeq
with
\begin{equation}
 L(\tau(e), e,d) = - \frac{1}{12} \tau_{bc}{}^a \tau_{ef}{}^d M_{ad} M^{be} M^{cf} - \frac{1}{4} \tau_{bc}{}^a \tau_{da}{}^c M^{bd} - 4 M^{ab} \nabla_a d \nabla_b d + 4 M^{ab} \nabla_a \nabla_b d \,. \label{parentDFT}
\end{equation}
After carrying out the Scherk-Schwarz reduction this becomes
\beq
S = v \int d \mathbb{X} \, e^{-2\hat{d}} \left( L(\bar{\tau}(\hat{e}), \hat{e}, \hat{d} ) + L_f ( \hat{e} , \hat{d}) \right)\,,
\eeq
where $v = \int e^{-2 \lambda} d\mathbb{Y}$. The first term in brackets is the same Lagrangian as before, except written in terms of the hatted vielbein $\hat{e}_{\mu}{}^A$ and the generalised torsion $\bar{\tau}_{BC}{}^A$ of the gauged DFT, as well as $\hat{d}$. The second term contains the dependence on the gaugings, and in terms of the torsion is given by
\beq
\begin{split}
L_f ( \hat{e} , \hat{d} ) = &
-\frac{1}{12} f_{BC}{}^A f_{EF}{}^D \hat{M}_{AD} \hat{M}^{BE} \hat{M}^{CF} - \frac{1}{4} f_{BC}{}^A f_{DA}{}^C \hat{M}^{BD} \\
 & -\frac{1}{6} \bar{\tau}_{BC}{}^A f_{EF}{}^D \hat{M}_{AD} \hat{M}^{BE} \hat{M}^{CF} - \frac{1}{2} \bar{\tau}_{BC}{}^A f_{DA}{}^C \hat{M}^{BD} \\
 & + 2 \hat{M}^{AB}\, \bar{\Gamma}^C_{AB} \, f_C  + 2 \hat{M}^{AB} \,\bar{\Gamma}^C_{CA} \, f_B \\
 & + 4 \hat{M}^{AB} \, f_A \, \partial_B \hat{d} \, - \hat{M}^{AB} f_A \, f_B \,.
\end{split}
\eeq
It is straightforward to check that this action exactly reproduces the gauged DFT action of \cite{Grana:2012rr}. 
Finally, it is important to mention that the Scherk-Schwarz action is only an $O_{D,D}$ scalar when one sets $f_A = 0$ \cite{Grana:2012rr}.

Due to the term \eqref{dEdE}, the parent DFT is only weakly invariant under local $H$-transformations, with
\begin{equation}
 \Delta_\lambda L(e,d) = Y^{ac}{}_{bd} e_{\nu}{}^b \partial_a e_\mu{}^d \partial_c \lambda^{\mu\nu} \, .
\end{equation}
We should check that the gauged DFT is also weakly invariant. To do so we note that by the Scherk-Schwarz ansatz, the $\mathbb{Y}$-dependence of the vielbein is introduced by the twist acting on the curved index
\begin{equation}
 e_\mu{}^a (\mathbb{X}, \mathbb{Y}) = \WI^a{}_A (\mathbb{Y}) \hat{e}_\mu{}^A (\mathbb{X})  \, .
\end{equation}
In order to preserve this after a Lorentz transformation, the transformations have to be solely $\mathbb{X}$-dependent, i.e. $\lambda^{\mu\nu} = \hat{\lambda}^{\mu \nu}(\mathbb{X})$. Then, the Lorentz variation of the action becomes
\begin{equation}
\begin{split}
 \Delta_\lambda L(e,d) &  = Y^{AC}{}_{BD} \left( \hat{e}_{\nu}{}^B \partial_A \hat{e}_\mu{}^D \partial_C \hat{\lambda}^{\mu\nu} -  \hat{e}_{\nu}{}^B \hat{e}_\mu{}^E \WI^{\! d}{}_{\! E} \WI^{\! a}{}_{\! A} \,\partial_a W^D{}_d \, \partial_C \hat{\lambda}^{\mu\nu} \right) \\ & \ssapprox 0 \, ,
 \end{split}
\end{equation}
and vanishes SS-weakly.

In summary, the approach of using the curvature-free but torsionful Weitzenb\"ock connection to construct the action works not only for the usual section condition but also when we allow a Scherk-Schwarz ansatz.

\section{Discussion}

In this paper, we discovered how to construct a geometry that gives the action of double field theory. For simplicity, we restricted ourselves to the NSNS sector of the closed string. The first step is to unify the fields into objects which are $O_{D,D}$ covariant. Also, the action is required to be invariant under generalised diffeomorphisms of the doubled spacetime. As a result one obtains the generalised metric, doubled dilaton and the generalised Lie derivative, which defines how objects transform under the action of generalised diffeomorphisms. This is a familiar path that has been discussed extensively in the literature \cite{Hull:2009mi,Hull:2009zb,Hohm:2010jy,Hohm:2010pp}.

What is new is our treatment of connections. One thing that is necessary if one is to do physics in a geometric setting is to have some notion of derivative that has well-defined transformation properties under diffeomorphisms. The information on how to achieve this is encoded in the connection. We found a connection that is metric-compatible, preserves the $O_{D,D}$ structure and maps generalised tensors into generalised tensors. The only necessary restriction is the use of a section condition that determines how the generalised spacetime is broken down into a physical spacetime, different solutions being related by T-duality. The connection that we find emerges naturally from the nonlinear realisation programme espoused in the work of \cite{Berman:2011jh}. 

Perhaps surprisingly, our connection has vanishing curvature. All of its non-trivial nature emerges from its torsion. In this sense, the Riemannian geometry of spacetime emerges from a more primitive structure based on teleparellelism, or in the language of Einstein \lq\lq Fernparallelismus'' \cite{Einstein:torsion}. General relativity can then be said to emerge at the point at which we choose to solve the section condition, leading to a physical spacetime embedded in the generalised spacetime.

The connection that we find is not invariant under local $H$-transformations and at first sight this might appear to be a big stumbling block. However, requiring local $H$-invariance at the level of the action as well as generalised diffeomorphism invariance renders the action unique.

We then apply our ideas to finding vacua of string theory that have a description as gauged supergravity. We do this by an analogue of the Scherk-Schwarz mechanism applicable to generalised geometry \cite{Dall'Agata:2007sr,Aldazabal:2011nj,Geissbuhler:2011mx,Grana:2012rr,Berman:2012uy}.

There remain many open questions. Can one make this action supersymmetric? Can one use this approach to write the higher derivative corrections in string theory
in double field theory using some collection of terms of order four in torsion? How does one extend this to the exceptional geometries of M-theory? How does one extend this to the geometries associated with the heterotic string?

\section{Acknowledgements}
DSB is partially supported by STFC grant consolidated grant $ST/J000469/1$. CB is supported by the STFC, the Cambridge Home and EU Scholarship Scheme, St John's College and the Robert Gardiner Memorial Scholarship. EM is supported by the STFC and a Peterhouse Research Studentship. MJP is in part supported by the STFC rolling grant $STJ000434/1$. MJP would like to thank the Mitchell family foundation and Trinity College Cambridge for their support. MJP and DSB would like to thank the Cook's Branch Nature Conservancy for its hospitality. DSB would like to thank Martin Cederwall and Daniel Thompson for discussions on various aspects of doubled geometry.

\appendix
\section{Weighted scalars and the dilaton} \label{Weighted}

\subsection{Transformation properties of weighted scalars}

Another part of the geometric picture which we must discuss is the definition of covariant derivatives of weighted scalars. This is necessary in order to include the dilaton in the double field theory.

Suppose that $S$ is a scalar of weight $w$, meaning its transformation law is
\beq
\delta_U S = \mathcal{L}_U S = U^c \partial_c S + w\, \partial_c U^c S\,.
\eeq
Then
\beq
\delta_U \partial_a  S = U^c \partial_a \partial_c S + \partial_a U^c \partial_c S + w \,\partial_c U^c \partial_a S + w\, \partial_a \partial_c U^c S \,.
\eeq
The first three terms are weakly equal to the generalised Lie derivative of a covector of weight $w$. The last term is anomalous and means that the partial derivative of a weighted scalar is not a weighted covector. However we can define a covariant derivative which is weakly a weighted covector by
\beq
\nabla_a S = \partial_a S  - w  \,\Gamma^b_{ba} S\,,
\eeq
as the anomalous terms in the transformation of $\Gamma^b_{ba}$ are weakly equal to $\partial_a \partial_b U^b$.

Now the dilaton $\Phi$ of string theory is incorporated into the double field theory by defining
\beq
e^{-2d} = e^{-2\Phi} \sqrt{|g|}
\eeq
to be a scalar of weight $w=1$ under generalised diffeomorphisms. The double field theory dilaton $d$ then acquires the transformation law
\beq
\delta_U d = U^c \partial_c d - \frac{1}{2} \partial_c U^c
\label{eq:dtransformation}
\eeq
and its covariant derivative is
\beq
\nabla_a d = \partial_a d  + \frac{1}{2} \Gamma^b_{ba} \,.
\label{eq:nablad}
\eeq
An unusual - but helpful - feature of this covariant derivative is that even though $d$ is not itself a scalar or a weighted scalar, $\nabla_a d$ as defined in \eqref{eq:nablad} is in fact (weakly) a covector.

\subsection{Dilaton in gauged DFT}

The Scherk-Schwarz reduction of the dilaton takes the form $d(\mathbb{X}, \mathbb{Y}) = \hat{d}(\mathbb{X}) + \lambda(\mathbb{Y})$. The dilaton transformation law in the gauged theory works out as
\beq
\hat{\delta}_{\hat{U}} \hat{d} = \hat{U}^A \partial_A \hat{d} - \frac{1}{2} \partial_A \hat{U}^A - \frac{1}{2} f_A \hat{U}^A \, ,
\eeq
where $f_A = \partial_a (W^{-1})^a{}_A - 2 ( W^{-1} )^a{}_A \partial_a \lambda$.

Note that the final term is clearly incompatible with defining the covariant derivative of $\hat{d}$ as $\hat{\nabla}_A \hat{d} = \partial_{A} \hat{d} + \frac{1}{2} \bar{\Gamma}^B_{BA}$, as we did in \eqref{eq:NablaHatdHat}, due to the fact the transformation law of the connection cannot involve $f_A$ (and would be needed to so as to cancel an anomalous term involving $f_B \partial_A \hat{U}^B$). This is another reason leading us to set $f_A = 0$.

\bibliographystyle{JHEP}
\bibliography{BibliographyL}

\end{document}